\documentclass[11pt,twoside]{article}
\usepackage{newpasp,epsf}
\def\edcomment#1{\iffalse\marginpar{\raggedright\sl#1\/}\else\relax\fi}
\reversemarginpar

\begin{document}
\title{Cosmic Particle Acceleration: Basic Issues}
\author{T. W. Jones}
\affil{University of Minnesota, Department of Astronomy, 116 Church St. SE,
Minneapolis, MN 55455, USA}

\begin{abstract}
Cosmic-rays are ubiquitous, but their origins are surprisingly difficult
to understand.
A review is presented of some of the basic issues common to cosmic 
particle accelerators
and arguments leading to the likely importance of diffusive shock
acceleration as a general explanation. The basic theory of diffusive
shock acceleration is outlined, followed by a discussion of some of 
the key issues that still prevent us from a full understanding of
its outcomes. Some recent insights are mentioned at the end that may help
direct ultimate resolution of our uncertainties.
\end{abstract}

\section{Introduction}
The inherent difficulty in understanding the acceleration of cosmic-rays
(CRs) may not immediately be obvious. At the most basic level we must presumably
identify an electric field capable of producing particles of very high 
energy. That sounds straightforward in fast moving plasmas.
For galactic and especially for ultra-high energy CRs, the energies
involved are so large that
the possibilities are very limited. When we consider, in addition,
the energy distribution of the CRs, as well as their composition,
rate of production and other details, however, the task of modeling
their production and propagation becomes very sophisticated. In this talk
I will deal mostly with a few of the more common and basic issues as they apply
to baryonic galactic CRs below the ``knee'', which we can conveniently
take to be $\sim 10^{6.5}$GeV/nucleus. Several speakers at this meeting have
admirably addressed many of the special issues relevant to other 
aspects of the broader problem. 

There is now broad consensus
that galactic CRs are accelerated mostly from the interstellar medium (ISM)
at supernova remnant blast waves by the diffusive shock acceleration (DSA) process. 
Beyond that simple statement, however, significant differences of
opinion quickly surface on almost every detail. Despite decades of
concerted and highly productive effort, this is not yet a 
solved problem, either physically
nor astrophysically.  I will now briefly outline some of the arguments
pointing us to the consensus viewpoint for the basic scenario, then
follow with a brief outline of a few of the issues that continue to hinder
our efforts to solve the problem fully. While these complicating issues
seem to be major barriers to a comprehensive understanding, there are
hints that when all the pieces of DSA theory
are in place together, a robust and possibly simple product may result. 
For additional DSA insights I direct readers to the accompanying discussion by Kang (2001),
which focuses on empirical evidence for DSA
as well as some of important numerical and 
technical issues and how they are being addressed.

\section{Background Issues}
Ultimately, CR acceleration comes through an electric field; 
however, the most convenient descriptions may not show this explicitly
for a given process.
The electric fields are most likely inductive, through
large scale motions, although they may be directly applied
through stimulated plasma waves. In any case we can express their
effective magnitude as ${\mathcal E} \sim \beta_a B$, where $\beta_a$ is the relevant
speed in the accelerator and $B$ is the strength of the local magnetic field.
For an accelerator of length scale, $R_a$, we can use this to constrain
the necessary magnetic field as
\begin{equation}
B > 10^{-5} {{E_7}\over{\beta_a Z~R_a(pc)}} {\rm Gauss},
\label{bscale}
\end{equation}
where $E_7$ is the required particle energy in units of $10^7$GeV and
$Z$ is the charge on the particle. This simple constraint can also
be derived through a number of different conceptual approaches
with modest variations in the numerical constant and some
variation in the interpretation of $\beta_a$. Examples include using
equation [5] for the time needed for DSA to
produce the required energies, constraining the diffusive length scale
of a particle
to be smaller than the size of the accelerator, or even just requiring
the particle gyroradius to be smaller than the size of the accelerator.

\begin{figure}
\plotone{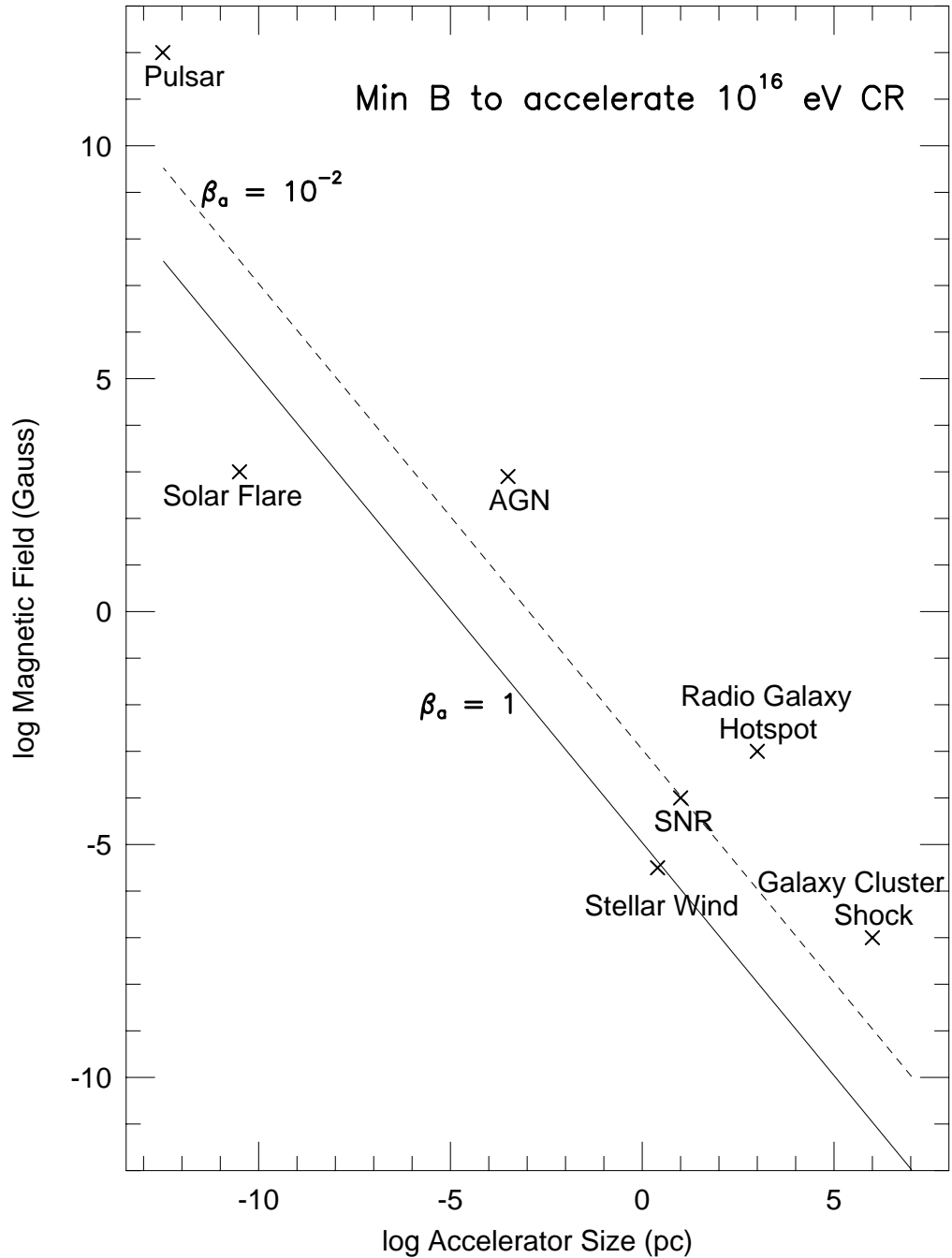}
\caption{Plot after Hillas (1984) showing a general constraint between
the size of a particle accelerator, its characteristic magnetic field
and the maximum energy possible in the accelerator. The diagonal lines
are lower bounds identified in relation [1] for a proton energy of $10^{16}$eV
and two different characteristic velocities in the accelerator. Rough
properties are estimated for several types of accelerators.}
\label{fig1}
\end{figure}
Figure 1 illustrates the result of equation 1 for $E_7 = 1$ and two values of 
$\beta_a$ in a form
made popular by Hillas (1984). In the figure I have indicated rough
conventional model properties for a small sample of astrophysical objects.
The only galactic objects known that may be able to satisfy the constraint are
supernova remnant shocks (``SNRs''), winds from O and B stars, pulsar
magnetospheres and possibly compact accreting binaries. 

Pulsars can be excluded as the principal source of galactic CRs
by considering CR composition. As described in detail by
other speakers at this workshop, the CRs below the ``knee'' roughly mirror the composition of the
sun and the ISM (e.g., Seo 2001; Wiebel-Sooth, Biermann \& Meyer 1998). 
There are important differences, including at the isotopic
level, that provide vital clues about the details of the source
plasma and CR propagation history (e.g., Meyer, Drury \& Ellison 1997). 
But, to lowest order this
information tells us that the source material is almost surely the ISM,
with perhaps some small admixture of locally processed stellar material.

Beginning from that point most models of galactic CR acceleration have 
focused on SNRs based on the total energy input required. That
can be estimated by considering the rate at which CR accelerators must 
replace CRs that diffuse from the galaxy. Isotopic
ratios fix the characteristic escape time near 100 MeV to be $\sim 10^7$yrs (Connell 1998).
Taking the observed local flux at the solar system and assuming this is
also the average for the galaxy this leads to
an average required CR power input near $10^{41}$ erg/sec 
(e.g., Drury, Markiewicz \& V\"olk 1989; Fields et al. 2000).  
If we take for comparison the
galactic supernova rate to be $\frac{1}{30}{\rm yr}^{-1}$, and kinetic
energy yield per event to be $10^{51}$erg, the available power in SNRs
is rough $10^{42}$ erg/sec, which is sufficient, but not by a
huge factor. Thus, SNR-based models must be at least moderately
efficient to supply the needed power. No other known galactic
source comes closer than an order of magnitude to this power supply,
explaining why most models have involved SNRs. 

The CR flux energy spectrum below the knee measured at earth is a 
power-law after correction for solar modulation,
\begin{equation}
\phi(E) \propto p^2f(p)\propto p^{-2.7}\propto E^{-2.7},
\end{equation}
where $\phi$ is the energy flux, and $f(p)$ is the phase space distribution
function (e.g., Seo 2001; Wiebel-Sooth et al. 1998). These scalings are strictly valid only when the CRs are relativistic.
In addition, the flux, $\phi$ is very nearly isotropic, consequent to
the diffusive propagation of CRs through the ISM. Propagation models
also generally lead to a steepening of the spectrum with respect
to its form at the source, by an increment $\sim 0.5-0.6$ in the
slope (e.g., DuVernois, Simpson \& Thayer 1996), reflecting
an energy dependence to the apparent CR escape rate. Thus, CR source models
usually aim to explain a power-law distribution function, $f(p) \propto p^{-q}$,
with $q \approx 4.1-4.2$. The fact that the simple steady state test particle
DSA theory predicts a power-law
$f(p)$ with $q \rightarrow 4+$ when shocks are strong is one of the primary reasons that
model for CR acceleration has attracted so much attention over the past
two decades. I will, in fact, limit my remaining discussion to issues
associated with this process.

\section{An Outline of Diffusive Shock Acceleration Theory}
There are a variety of approaches to understanding the physics underlying
DSA, since the microphysics is 
complex and depends on what one assumes about such details as the
structure and orientation of the local magnetic field. All approaches
depend on a small fraction of nonthermal particles becoming trapped by scattering
around a shock front, so that they may tap into the energy
flow through the shock for extended times, but with a finite probability of
escaping in a given time interval. 
It is remarkable that all these approaches give virtually the same answer
to the first approximation, at least so long as
feedback on the plasma flow can be neglected and various pathologies
are avoided. For in-depth discussions of the theory there are a number of fine
reviews (e.g., Berezhko \& Krymskii 1988; Blandford \& Eichler 1986;
Drury 1983; Jones \& Ellison 1991; Malkov \& Drury 2000). Here I will outline only some basic properties of
the theory as it applies to quasi-parallel shocks, since that is
one of the simplest situations.

The DSA theory depends on an almost isotropic particle distribution and
particle speeds large compared to the bulk flow speed, $u$.
Spatial gradients limited by the assumption of diffusive propagation
with respect to local scattering centers lead to a transport
equation of the diffusion-convection type (e.g., Parker 1965; Skilling 1975a),
\begin{equation}
\frac{\partial f}{\partial t} + u\cdot\nabla f =- \frac{1}{3}p\frac{\partial f}{\partial p}
\left(\nabla \cdot u\right) + \nabla\cdot\left(\kappa\nabla f\right)
+ Q,
\end{equation}
where the first term on the right accounts for adiabatic compression,
the second spatial diffusion and $Q$ is a generic source term that
can represent injection or escape, for example. The diffusion coefficient
is given by $\kappa = \frac{1}{3} \lambda v$, where $\lambda$ is the
scattering length, and $v$ is the particle speed. The flow velocity
is u. In this context the
scattering is commonly assumed to involve resonant Alfv\'en waves, 
whence one can estimate from quasi-linear theory (e.g., Skilling 1975b) a scattering length 
$\lambda = \zeta r_g$, with $r_g$ the particle gyroradius and
\begin{equation}
\zeta = \frac{4}{\pi}\frac{P_B}{k P_{wk}} = \frac{4 P_B}{\pi P_w} \sim \frac{B^2}{(\delta B(k))^2}.
\end{equation}
Here $P_B$ is the total magnetic pressure and $k P_{wk}$ is the 
pressure (energy density) in waves satisfying resonance,
which can be ``sharpened'' to be expressed as $kp = \omega_cm$, or
$k r_g = 1$,
where $\omega_c$ is the nonrelativistic cyclotron frequency for
the particle species under discussion. The limiting value, $\zeta = 1$
($\lambda = r_g$) leads to so-called Bohm diffusion.
For an oblique magnetic field at a plane shock this same formalism
applies with the substitutions $\kappa \rightarrow \kappa_{\parallel}$, then
$\kappa = \kappa_{\parallel} cos^2{\theta} + \kappa_{\perp} sin^2{\theta},$
with $\theta$ the angle between the magnetic field and the shock normal
and $\kappa_{\perp} = \kappa_{\parallel}/(1 + \zeta^2)$ (e.g., Jokipii 1987).
Then $\kappa$ refers to diffusion along the shock normal, while
$\kappa_{\parallel}$ and $\kappa_{\perp}$ describe diffusion along and
perpendicular to the local mean magnetic field.

For a parallel shock ($\theta = 0$) we can imagine the acceleration as a first order
Fermi process, with particles successively being scattered across
the shock from opposite sides of a converging flow. 
The mean fractional momentum gain between successive
downstream returns is $\frac{\Delta p}{p} \approx \frac{4|\Delta u|}{3 v}$,
where $\Delta u = u_2 - u_1$ is the velocity change across the shock.
The probability of downstream escape by advection following each
downstream return is simply $P_{esc} \approx 4u_2/v$. Since the
average distance a particle diffuses on either side of the shock before
being returned is $x_{d_{1,2}} = \kappa_{1,2}/u_{1,2}$, the mean time
between crossings is $t_{sc} = x_d/v$.
Using this one can estimate the mean time for a particle
to be accelerated from $p_1$ to $p_2$ as  (e.g., Lagage \& Cesarsky 1983)
\begin{equation}
t_a = \frac{3}{|\Delta u|}\int^{p_2}_{p_1} \left[\frac{\kappa_1}{u_1}+\frac{\kappa_2}{u_2}\right] \frac{dp}{p}\\
\rightarrow ({\rm factor})\times \frac{x_{d_1}(p_2)}{u_1}=({\rm factor})\times t_d(p_2),
\end{equation}
where the arrow represents a trend to the Bohm limit, and the
``factor'' is 20 for a limiting strong gas shock, if $\kappa_1 = \kappa_2$,
and $p_2 >> p_1$. Requiring $t_a <  R/u_1$, we recover relation [1] multiplied
by a factor $\frac{20}{3}$, with $\beta_a = \frac{u_1}{c}$ in the Bohm limit. Note that
$x_d/u_1 = t_d$, called the diffusion time for the CRs, is 
the time scale over which an isotropic population of CRs will be
advected across the diffusive ``precursor'' formed ahead of a shock.
This is also the average time for CRs to diffuse a length $x_d = \kappa/u_1$.

For a plane gas shock the steady
solution to equation [3] is a power-law with 
\begin{equation}
q = \frac{3u_1}{u_1-u_2} = \frac{3r}{r-1}\rightarrow\frac{4M^2}{M^2+3},
\end{equation}
where $r = \frac{u_1}{u_2} = \frac{\rho_2}{\rho_1}$ is the compression
ratio of the shock, $M$ is the shock Mach number, 
and the arrow corresponds to  a $\gamma = \frac{5}{3}$ gas.
Then $q \rightarrow 4$ in the strong shock limit. 
A simple computation shows that $q = 3 + P_{esc}\times \frac{p}{\Delta p}$,
so kinematically the spectrum reflects the match between the rate of
particle acceleration and escape.
This solution neglects
any backreaction of CRs on the shock structure, so constitutes a
``test particle'' solution to the problem.

That simple, limiting solution, apparently independent of any
microphysical details, and naturally leading very close to the expected 
source slope for galactic CRs was one of the key insights that raised the
community's consciousness about DSA in the
late 1970s 
(Axford, Leer \& Skadron 1977; Krymskii 1977; Bell 1978; Blandford \& Ostriker 1978). 
The other was a realization that the Alfv\'en
waves needed to isotropize the CRs would be generated by the CRs
themselves as they streamed into the oncoming upstream plasma.
The postshock plasma is expected to be turbulent (e.g., Quest 1988), including
waves advected from upstream, so scattering in that region seemed
assured, as well.
Quasilinear theory provides an estimate of the growth time
for resonant Alfv\'en waves ahead of the shock, which depends on $\nabla f$.
Using the fact that the CRs will diffuse upstream a characteristic length,
$x_d$, so that $\nabla f \propto f(p)/x_d$, we can estimate this time to be
\begin{equation}
t_w \sim \frac{x_d}{v_A}\frac{P_w}{P_c(p)} \sim t_d \frac{M_A}{\zeta} \frac{P_B}{P_c(p)}.
\end{equation}
Here $P_c(p)$ is the pressure in resonant CRs, $v_A$ is the upstream Alfv\'en
speed, and $M_A$ is the Alfv\'enic Mach number of the shock.
Thus, on the surface, this time scales with the acceleration time, $t_a$, although
the additional factors, which are not all known or even constant, 
complicate the comparison considerably. In practice
calculations have generally assumed $t_w$ is very short, so that
$P_w$ reaches an asymptotic limit (commonly given as Bohm diffusion), or 
that wave dissipation and growth are in a local equilibrium that
produces another preferred diffusion coefficient (e.g., Jones 1993).
In fact, except for some hybrid plasma simulations involving a limited
range of particle energies (e.g., Quest 1988; Ellison, et al. 1993) this is not a solved problem.

\section{Some Important Details}
The real beauty of DSA was its apparent simplicity and the robust
character of the solution. However,
virtually as it was introduced, DSA theory
exposed some potentially very messy details that make the simple
behaviors outlined above not obviously valid in many applications.
One I will mention but not elaborate here is the influence of an
oblique magnetic field at the shock. Then the magnetic field
component aligned with the shock face jumps at the shock, and that
means there is necessarily an electric field also aligned with the shock
face, but orthogonal to the magnetic field. This leads to ``shock drift
acceleration'' and additional possible complications if the particle propagation
is not diffusive across field lines (e.g., Gieseler et al. 1999). 
Jokipii (1982) showed, however, that so long
as the particles diffuse across field lines and the shock is planar, 
then the basic formalism already outlined remains intact even for 
$\theta \sim \pi/2$.  When $\zeta >> 1$, however,
the length and time scales for highly oblique shocks can be
reduced from Bohm diffusion values (Jokipii 1987). Another complication can result if the
magnetic field itself ``wanders'' or is ``braided'' (Kirk, Duffy \& Gallant 1996), since particle
motions along the field lines can lead to additional spatial transport
that influences the rates of acceleration and escape. While
some shocks may well be almost perpendicular, with $\theta \sim \pi/2$,
it seems to me if the magnetic field is moderately turbulent,
that global perpendicular shock behaviors are not likely to be prevalent
for nonrelativistic shocks.

Many of the other complications can be summarized by noting that
DSA is an integral part of collisionless
shock formation itself (Eichler 1979). The particles we call CRs are really just a
nonthermal tail of the full distribution, $f(p)$.
They are distinguished by their relatively long scattering lengths,
and, of course, the relatively large individual particle energies,
reflecting the absence of full thermodynamical equilibrium.
The  CR particles
come from, or ``are injected'' from, the more abundant bulk plasma
population. They also exert a pressure through their interactions
with Alfv\'en waves that can modify the flow of the bulk plasma.
The strength of those waves depends, in turn, on the intensity of the CR
streaming, as already pointed out, and on comparative wave
growth and dissipation rates. Wave dissipation heats the plasma
as well.
Flow modifications from a classical gas shock structure resulting from
these features alter the various terms in the diffusion-convection 
equation [3], so a test particle solution based on an unmodified
shock flow must be reexamined in light of a more integrated view
of the physics.

\subsection{Dynamical Backreaction}
The possible importance of CR backreaction was quickly recognized
from estimates of the likely CR pressures (Axford et al. 1977; Eichler 1979). That can be
expressed as 
\begin{equation}
P_c = \frac{4\pi}{3} \int^{p_2}_{p_1} v p^4 f(p) d\ln p = \int^{p_2}_{p_1} P_c(p) d \ln p.
\end{equation}
For $f(p) \propto p^{-q}$ this diverges logarithmically as $q \rightarrow 4$  for
strong shocks and as
$p_2 \rightarrow \infty$. Clearly a real shock must
at least include a cutoff at finite $p_2$. 
Such a cutoff appears naturally from equation [5] in a shock of
finite age, or as a result of escape by CRs above some momentum
The latter effect may result either as a consequence of finite shock
extent, or from limitations to the intensities of Alfv\'en waves
resonant with the highest momentum particles. I will revisit this last
point later on.

Using equation [7] we can crudely compare $P_c$ at the shock for a given CR number
density to the thermal pressure, $P_t = n_t k_bT \sim n_t p_t v_t$,
where $p_t = m v_t$ represents the (nonrelativistic) momentum of a
thermal ion. Taking $q = 4$ to illustrate, and assuming for simplicity
that the CRs are all relativistic, we have
\begin{equation}
\frac{P_c}{P_t} \sim \frac{n_c}{n_t}\frac{p_1 c}{p_t v_t} \ln{\frac{p_2}{p_1}}.
\end{equation}
Since all the terms on the right are large except for the fraction of
ions in the CR populations, $n_c/n_t$,
it becomes immediately obvious when $p_2/p_1 >> 1$ that
a relatively very small CR population can easily produce a pressure 
comparable to the thermal gas.

The consequences of backreaction to the shock properties are significant.
First, the gradient from a finite $P_c$ slows and
compresses plasma adiabatically before it reaches the classical
and discontinuous gas ``subshock''.
This effect can be simply estimated 
by recalling that the CRs are distributed upstream in a precursor
of characteristic length $x_d$, producing a pressure gradient
that decelerates the flow as it approaches the shock by an amount
\begin{equation}
\frac{\Delta u_1}{u_1} \approx \frac{\partial P_c}{\partial x}
\frac{1}{\rho_1}\frac{x_d}{u_1} \frac{1}{u_1} \approx \frac{P_c}{x_d}\frac{x_d}{\rho_1 u^2_1} = \frac{P_c}{\rho_1 u^2_1}.
\end{equation}
As $P_c$ at the shock becomes comparable to the dynamical momentum flux into the
shock, $\rho_1u^2_1$, we expect the subshock to become very much weakened, since
$\Delta u_1/u_1 \sim 1$, and adiabatic heating in the shock precursor
will reduce the Mach number of the flow entering the subshock.

The first and simplest approach to evaluating in detail the dynamics of modified
CR shocks used the energy moment of the diffusion-convection  equation (Drury \& V\"olk 1981; Axford, Leer \& McKenzie 1982).
Defining
\begin{equation}
E_c = 4\pi mc^2 \int^{p_2}_{p_1} p^3 \left(\sqrt{p^2+1}-1\right)f(p)d\ln{p},
\end{equation}
and the closure relation, $P_c \equiv (\gamma_c - 1)E_c$, we have from equation [3]
\begin{equation}
\frac{\partial E_c}{\partial t} + u\cdot\nabla E_c = -\gamma_c\nabla\cdot u
+ \nabla\cdot\left(<\kappa >\nabla E_c\right) + S,
\end{equation}
provided the limits $p_1$ and $p_2$ can be neglected.
The term $<\kappa>$ is a mean diffusion coefficient weighted by momentum
and $f(p)$,
and $S$ is an integral form of the source term, $Q$, in equation [3].
In equation [11] I have expressed $p$ in units of $mc$.
When merged with Euler's equations for gas dynamics, this approach is
commonly termed a ``two-fluid'' dynamical model, since the CRs are
treated as a massless, diffusive fluid coupled to the bulk flow.
Backreaction on the bulk plasma is included through the pressure gradient
terms in the bulk flow momentum and energy equations. The 
pondermotive force of the Alfv\'en waves, as well as their energy dissipation
may also be readily included (Achterberg 1982; McKenzie \& V\"olk 1982). 
The two-fluid approach is somewhat
controversial, mostly because the closure parameters, $\gamma_c$ and
$<\kappa>$ are not known a priori, and because of some pathological
steady state solutions for strong shocks in the limit $\gamma_c = \frac{4}{3}$.
Nonetheless, when properly used, it is an effective and computationally
efficient method to establish basic features of modified CR shocks (Kang \& Jones 1995).

The first two-fluid computations showed that it was possible for
most of the momentum flux through a shock to be converted into
CR pressure, for example, with $P_c$ amplified over the
precursor length, $x_d$.
As many as three steady solutions were identified for strong shocks
from given upstream conditions or particle injection rates,
when $\gamma_c \approx \frac{4}{3}$ (Drury \& V\"olk 1981). 
The solutions differ substantially in the ``efficiency'' of
conversion by the shock of momentum influx, $\rho_1u_1^2$, into $P_c$,
and represent a bifurcation phenomenon with respect to the supply of
seed particles.
The pathological steady state solutions involved finite postshock 
$P_c$ from zero upstream $P_c$, and completely smoothed shocks, in 
which the gas subshock disappears. Those solutions are the result of
assuming $p_2 \rightarrow \infty$ and cannot be reached practically
from time
dependent solutions, or when $p_2$ is finite. Nonetheless, as I will
outline below from other considerations, the discovery from two-fluid
models is correct that shocks may either be highly efficient or
not very efficient in accelerating CRs in ways that depend
sensitively on the supply of seed CRs.
Two-fluid calculations also confirmed that the basic
evolutionary timescale for shock modification is $t_d = x_d/u_1$
(Drury \& Falle 1986), and identified the existence of dynamical instabilities
derived from the long scaled coupling between the bulk plasma and
the CRs (Drury \& Falle 1986; Zank, Axford \& McKenzie 1990; Ryu, Kang \& Jones 1993).

\subsection{Nonlinear Modifications to the CR Spectra}
According to the rightmost expression of equation [6] the CR spectrum should steepen 
when a pressure precursor forms at the shock, so that the Mach number of
the subshock is reduced by adiabatic heating and deceleration of the
inflow.
That is a somewhat misleading observation, however, since the total compression
across the structure, including the precursor, is greater
than for the gas shock alone. Thus, according to the other expressions
for $q$ in equation [6] particles scattered across the full shock
transition, where $r > 4$ for an initially strong shock, 
would be expected to form into a very hard spectrum with $q < 4$.
The relevant interaction length for the CRs is, $x_d$, of course.
For Bohm-like diffusion, with $\kappa \propto pv$, we have $x_d(p) \propto pv$,
so particles at relatively low momenta respond mostly to the jump across
the gas subshock, while the highest momenta particles will reach well past
the diffusion lengths of those lower momentum particles. From relations
[9] and [10] it is clear that modest momentum CRs may produce significant compression 
in front of the subshock. That property tends to produce concavity in the form of $f(p)$,
and that, in turn enhances the relative importance to $P_c$ of the highest
momentum particles over the expression [9]; that is, the divergence of
the pressure is faster than logarithmic with $p_2$.
Thus, the hydrodynamical form of the shock and the form of the CR 
distribution, $f(p)$, are linked in a highly nonlinear manner.
This point is crucial to our understanding of DSA
in practice, as pointed out by a number of authors (e.g., Ellison \& Eichler 1984; Malkov 1999).
The development of these nonlinear features in such a modified CR shock is 
clearly visible in the evolution of the fully nonlinear diffusion-convection-equation-based 
simulation presented in (Kang 2001; Kang et al. 2001), for example.

It is not yet entirely clear what a strongly modified CR shock will look
like when it is examined in a complete and fully self-consistent way, 
nor what the CR spectrum is, despite a considerable effort
put into determining those issues. I will return at the end to some
recent insights into those questions. Before that it is useful to
complete our discussion of issues with a few comments on two more critical
aspects of the problem; namely, the injection of CRs out of the bulk
plasma, and feedback between the CR acceleration and the evolution
of the Alfv\'enic turbulence responsible for moderating the acceleration.

\subsection{The Critical Role of Injection}
Several authors have pointed out that the efficiency of CR acceleration
at strong shocks depends sensitively on the rate of injection there (e.g., Eichler 1979; Berezhko et al. 1995; Malkov 1999).
Berezhko et al. (1995)  argued for the existence of a critical
injection rate, above which the shocks are highly efficient, so that
$P_c$ at the shock is a large fraction of the momentum flux into the shock.
Below such a threshold the process becomes much less efficient, 
so the pattern is reminiscent of the original two-fluid results.
As part of a study of CR acceleration in SNRs Berezhko et al. found
a sharp increase in acceleration efficiency in high Mach number shocks
as the injected proton fraction was increased between $10^{-4}$ and $10^{-3}$
of the number flux through the shocks. They arbitrarily fixed that
number in the models, but the simulations still highlight the issue clearly.
Malkov (1997a,b) also argued for an injection threshold on the 
grounds that once a shock
begins to be modified, so that the compression is increased, the highest
momentum particles see a larger velocity jump, so are more effectively 
accelerated, thus enhancing $P_c$. That, in turn, enhances the
compression at the shock, increasing the acceleration rate, and so on. 
The process then becomes limited by
the highest momentum to which CRs can be accelerated.

It becomes crucial, therefore, to incorporate appropriate injection physics
in DSA models.
As noted, CRs are an extension of the thermal particle pool reflecting
the absence of full equilibrium. At a shock the majority of ions are
``thermalized'' and unable to re-cross the shock, since the shock thickness
is determined by the characteristic thermalization length of the ions. That
process is very complex and incomplete in a collisionless plasma (Kennel,
Edmiston \& Hada 1985), 
however, and some fraction of ions having been only partially thermalized
may escape upstream as ``seed particles'' for DSA.
The injection problem amounts to determining how that seeding, or ``thermal
leakage'' is controlled. Monte Carlo simulations handle the process
very simply by assigning a form to the scattering law for all the ions
that allows a smoothly increasing escape probability with increasing
momentum; i.e., by setting $\lambda \propto p^\beta_s$ with $\beta_s > 0$ (e.g., Baring, Ellison \& Jones 1994).
While that captures the flavor of the process, it does not attempt to 
include an explicit model for the nonlinear plasma physics associated
with the thermalization process. Hybrid plasma simulations do that
in detail, of course, but are not really designed to explore the
production of very high energy particles that may be accelerated at
cosmic shocks. 

Malkov (Malkov \& V\"olk 1995; Malkov 1998) has recently developed a very promising analytical
model for thermalization and associated injection based
on the nonlinear trapping of ions in Alfv\'en waves generated by
ions escaping upstream and amplified through the shock. This model
is calibrated against hybrid plasma simulations, so contains no 
free parameters. They find a very
sharp cutoff in the probability to return upstream.
For strong shocks only ions with speeds more than roughly 10 times the bulk flow speed
away from the shock have a finite chance to escape back into the oncoming
flow and become seed CRs. That is a pretty strong filter, and in
nonlinear diffusion-convection  simulations we carried out recently
based on this model (Gieseler, Jones \& Kang 2001), the injected proton fraction quickly stabilized
around $10^{-3}$. For this model that result seems fairly robust,
with an equilibrium formed between reduced injection coming from
cooling of postshock gas as seed particles escape, against
decreased upstream compression and a stronger subshock resulting from 
reduced $P_c$ if injection falls below the equilibrium value.

\subsection{Alfv\'en Wave Feedback Loops}
By now it should be clear how intricately connected the different
elements of the nonlinear DSA model are. Once injected
the rate at which particles are accelerated in a parallel shock
depends on two things; namely,
the velocity profile of the bulk flow and the spectrum and intensity
of resonant Alfv\'en waves across the flow profile. As mentioned at the beginning, 
amplification of the Alfv\'en waves
in the precursor is generally attributed to instabilities fed by the CRs
themselves as they attempt to stream ahead of the shock. While the
quasilinear theory of wave amplification is well-established, and
was used in deriving equation [7], for example, once the wave amplitudes
become large, quasi-linear theory is suspect. Similarly, wave
dissipation is generally attributed to nonlinear Landau damping (e.g., V\"olk, Drury \& McKenzie 1984),
but again that has not been developed to a state that it can be reliably
used to give an accurate, fully nonlinear treatment of source
terms for the Alfv\'enic turbulence.
The wave dissipation is important in another way, since it leads to
local heating of the bulk plasma, adding to the adiabatic heating
that already reduces the Mach number at the gas subshock. This heating acts as
another limiter in the acceleration process.

Malkov, Diamond \& Jones (2001) have also pointed out
an important detail in the transport of Alfv\'enic
turbulence in modified shocks that strongly
influences the maximum particle momentum, $p_2$, that, we will
recall, becomes the controlling influence in the efficiency of
DSA at modified shocks. 
To see this, consider at a given time
that the maximum momentum for the CRs is $p_2(t)$, and then follow
those particles as they subsequently return from the downstream
flow moving into the upstream flow
with increased momenta, $\tilde p_2 = p_2(t) + \Delta p$, after scattering. Being the first to stream into the
flow with these momenta, they do not encounter significant Alfv\'en
wave amplitudes at resonant wavelengths ($kr_g \sim 1$). Thus, they
should easily escape the shock and will not be further accelerated. Their streaming
will, on the other hand, amplify whatever low level Alfv\'en waves are upstream
at the resonant wavelength, and those waves will be advected towards the
shock where they can interact with subsequent CRs traveling upstream.
However, in a strongly modified shock the flow is compressed as it
approaches the gas subshock, causing the in-flowing Alfv\'en wave
to be compressed, as well. (The Alfv\'en speed will generally decrease.)
Therefore, these waves will be resonant only with CRs of smaller momentum than
$\tilde p_2$, and the current population of CRs at $\tilde p$
will not be scattered until they propagate
to regions where there has been no compression to the flow. 
Their rate of return is consequently reduced by the preshock compression,
so that there may be a substantial reduction in $p_2(t)$
from the one predicted from Bohm diffusion.

\section{Discussion: Resolving these Issues}
The preceding section may leave one with less than full confidence that
we will soon be able to model fully nonlinear modified CR shocks
in a complete and self-consistent manner. There are, however, some
encouraging developments that could lead us into a much clearer
understanding. For one, computational techniques, as discussed by
Kang (2001), are advancing rapidly and hopefully will
soon allow us to include explicitly most or all of the physics outlined
here. Second, some recent insights suggest that the full solution for
modified shocks may turn out to be robust after all, and that all of 
the complications work together in a way to find a ``critical'' 
solution. 

For example, Malkov (1999) demonstrated recently the existence of a
similarity solution for a steady, strongly modified CR shock. He found a 
solution to the coupled diffusion-convection  and Euler's equations
in which the flow is highly modified, even allowing the
total compression to become arbitrarily large. Those were exactly the
kinds of situations where concerns were raised above about
the development of non-homologous spectra dependent on many details.
However, despite the highly nonlinear character of Malkov's similarity solution,
the CR distribution function
takes a simple power-law of the form, $f(p) \propto p^{-7/2}$,
independent of the total compression and the form of a given $\kappa(p) \propto p^{\beta_s}$,
so long as $\beta_s > 0.5$. In fact,
the flow profile adjusts to the given form of $\kappa$, so that the
particle distribution takes this ``universal form''. 
In that work Malkov additionally confirmed from the
diffusion-convection equation the existence of three distinct
solutions analogous to those mentioned earlier, giving very different 
results for the efficiency of the acceleration and consequent shock modification.
This calculation also established that no solution with a 
vanishingly small gas subshock can result from the diffusion-convection
equation.
Malkov identified a critical parameter $\frac{\eta p_2}{M^{3/4}}$
determining the character of the solution and established
for a given Mach number, $M$, and CR energy injection efficiency, $\eta$,
that the existence of test-particle and high efficiency solutions
depends on the value of $p_2$. Small values lead to the test-particle
solution, while large values formally open up ``intermediate
efficiency'' and the ``high efficiency'' solutions.

As an extension of this insight, Malkov, Diamond and V\"olk (2000)
recently argued that modified CR shocks may evolve towards a
{\it self-organized critical state} that finds the combination
of critical parameters balancing the energetics of the shock at
an appropriate equipartition among the components; i.e, an ``attractor
state.'' The critical solution for a given Mach number
depends on the self-adjustment of the particle injection rate and
the maximum momentum, $p_2$. Those in turn are coupled to each other
and to the growth, propagation and dissipation of the resonant
scattering waves and the underlying bulk plasma that complete the basic 
physical system. Those authors suggest that the solution would correspond to
the shock compression and injection rate that give exactly one solution.
That, in turn depends on the maximum particle momentum, $p_2$. Were
$p_2$ to increase above its critical value the total compression would
increase, weakening the subshock and reducing the injection rate. On the
other hand if increased escape rates reduced $p_2$ the total compression
would be expected to decrease, leading to an increase in the injection rate.
Both effects would serve to return the system to its critical point.

While these fascinating insights are yet to be confirmed by direct
simulations, they do remind us that DSA is one piece of the full physics
of collisionless shock formation, and encourages us to look for a unified
view of the results. Then perhaps we can anticipate a relatively
simple outcome to match the empirical result that CRs do seem to
appear widely with power-law spectra, implying, when viewed as a
whole, that the details are
not so crucial after all.
That ``whole'' is currently beyond our understanding, but we are making
good strides towards clarifying it.

\acknowledgments
This work has been supported by NASA through grant NAG5-5055 and by
the University of Minnesota Supercomputing Institute. I am grateful
to my colleagues for stimulating discussions on many of these topics, and
especially to Hyesung Kang and to Mischa Malkov for novel insights. 
Also, I thank the organizers of this meeting for their hard work and
for providing a provocative and illuminating program.


\begin{references}

\reference Achterberg, A. 1982, \aap, 98, 195

\reference Axford, W. I., Leer, E. \& McKenzie, J. F. 1982, \aap, 111, 317

\reference Axford, W. I., Leer, E. \& Skadron, G. 1977, Proc. 15th I.C.R.C.,
11, 132

\reference Baring, M. G., Ellison, D. C. \& Jones, F. C. 1994, \apjs, 90, 547

\reference Bell, A. R. 1978, \mnras, 182, 147

\reference Berezhko E.G., \& Ellison, D. C. 1999, \apj, 526, 385

\reference Berezhko, E. G. \& Krymskii, G. F. 1988, Soviet Phys. Usp., 31, 27

\reference Berezhko E.G., Ksenofontov L.T. \& Yelshin V.K., 1995,
Nuclear Phys. B, 39A, 171

\reference Blandford R.~D., \& Eichler D. 1987, Physics Reports, 154, 1

\reference Blandford, R. D. \& Ostriker, J. P. 1978, \apjl, 221, L29

\reference Connell, J. J. 1998, \apjl, 501, L59

\reference DuVernois, M. A., Simpson, J. A. \& Thayer, M. R. 1996, \aap, 316, 555

\reference Drury L.~O'C. 1983, Rep. Prog. Phys., 46, 973

\reference Drury, L. O'C. \& Falle, S. A. E. G. 1986, \mnras, 223, 353

\reference Drury, L. O'C., Markiewicz, W. J. \& V\"olk, H. J. 1989, \aap, 225, 179

\reference Drury, L. O'C. \& V\"olk, H. J. 1981, \apj, 248, 344

\reference Drury, L.~O'C., \& Falle, S.~A.~E.~G. 1986, \mnras, 223, 353

\reference Eichler, D. 1979, \apj, 229, 419

\reference Ellison, D. C. \& Eichler, D. 1984, \apj, 286, 691

\reference Ellison, D. C., Giacalone, J., Burgess, D. \& Schwartz, S. J. 1993,
J. Geophys. Res., 98, 21085

\reference Fields, B. D., Olive, K. A., Cass\'e M. \& Vagioni-Flam, E. 2000,
astro-ph/0010121

\reference Gieseler, U. D. J., Jones, T. W. \& Kang, H. K. 2001, \aap (in press)

\reference Gieseler, U. D. J., Kirk, J. G., Gallant, Y. A. \& Achterberg, A. 1999, \aap, 345, 298

\reference Hillas, A. M. 1984, \araa, 22, 425

\reference Jokipii, J.R. 1982, \apj, 255, 716

\reference Jokipii, J.R. 1987, \apj, 313, 842

\reference Jones, F. C. \& Ellison, D. C. 1991, Space Sci. Rev., 58, 259

\reference Jones, T. W. 1993, \apj, 413, 619

\reference Kang, H. 2001, this proceedings

\reference Kang H., \& Jones T.W. 1995, \apj, 447, 944

\reference Kang H., Jones T.W., LeVeque, R., \& Shyue, K. M. 2001, \apj,
March 20 issue, astro-ph/0011538

\reference Kennel, C. F., Edmiston, J. P. \& Hada, T. 1985, in
``Collisionless Shocks in the Heliosphere: A Tutorial Review'', ed.
R. Stone \& B. Tsurutani (Washington, D. C.: Amer. Geophys. Union), p 1
\reference Kirk, J. G., Duffy, P. \& Gallant, Y. A. 1996, \aap, 314, 1010

\reference Krymsky, G. F. 1977, Dokl. Akad. Nauk USSR, 234, 1306

\reference Lagage, P.O., \& Cesarsky, C.J. 1983, \aap, 118, 223

\reference Malkov, M. A. 1997a, \apj, 485, 638

\reference Malkov, M. A. 1997b, \apj, 491, 584

\reference Malkov, M. A. 1998, Phys. Rev. E, 58, 4911

\reference Malkov, M. A. 1999, \apjl, 511, L53

\reference Malkov, M. A., Diamond, P. H. \& V\"olk, H. J. 2000, \apjl, 533, L171

\reference Malkov, M. A., Diamond, P. H. \& Jones, T. W. 2001, in preparation

\reference Malkov, M. A. \& Drury, L. O.'C. 2000, Rep. Prog. Phys. (submitted)

\reference Malkov, M. A. \& V\"olk 1995, \aap, 300, 605

\reference McKenzie, J. F. \& V\"olk H. J. 1982, \apj, 116, 191

\reference Meyer, J.-P., Drury, L. O'C. \& Ellison, D. C. 1997, \apj, 487, 182

\reference Parker, E. N. 1965, Planet Space. Sci., 13, 9

\reference Ryu, D., Kang, H. \& Jones, T. W. 1993, \apj, 405, 199

\reference Quest, K.B. 1988, J. Geophys. Res.,  93, 9649

\reference Seo, E-S. 2001 (this proceedings)

\reference Skilling J. 1975a, \mnras, 172, 557

\reference Skilling J. 1975b, \mnras, 173, 255

\reference V\"olk, H. J., Drury, L. O'C. \& McKenzie, J. F., 1984, \aap, 130, 19

\reference Wiebel-Sooth, B., Biermann, P. L. \& Meyer, H. 1998, \aap, 330, 389

\reference Zank, G. P., Axford, W. I. \& McKenzie, J. F. 1990, \aap, 233, 275

\end{references}
\end{document}